\documentclass[10pt,prl,preprintnumbers,floatfix,aps,nofootinbib,showpacs,twocolumn,superscriptaddress]{revtex4} 

\usepackage{epsfig}
\usepackage{url}
\usepackage{hyperref}

\usepackage[normalem]{ulem} 

\usepackage{latexsym}
\usepackage{epsfig}
\usepackage{amsmath}
\usepackage{amssymb}
\usepackage{wasysym}
\usepackage{graphicx}
\usepackage{verbatim}
\usepackage{enumerate,mdwlist}
\usepackage[titletoc]{appendix}
\usepackage{amsfonts}
\usepackage{tikz} 

\usepackage[normalem]{ulem}

\newcommand{\lp}{\left(}
\newcommand{\rp}{\right)}

\newcommand{\ba}{\begin{eqnarray}}
\newcommand{\ea}{\end{eqnarray}}
\newcommand{\be}{\begin{equation}}
\newcommand{\ee}{\end{equation}}

\newcommand{\C}{\mathcal{C}}
\newcommand{\W}{\mathcal{W}}

\definecolor{grey}{rgb}{0.4,0.4,0.4}
\definecolor{dullmagenta}{rgb}{0.4,0,0.4}
\definecolor{darkblue}{rgb}{0,0,0.4}
\definecolor{midblue}{rgb}{0,0,0.5}
\definecolor{midred}{rgb}{0.5,0,0}
\definecolor{orange}{rgb}{1,0.5,0}
\definecolor{lightbrown}{rgb}{0.75,0.5,0.25}
\definecolor{tan}{cmyk}{0.14,0.42,0.56,0}
\definecolor{djunglegreen}{cmyk}{0.99,0,0.52,0}
\definecolor{lightgreen}{rgb}{0,1,0}
\definecolor{olivegreen}{cmyk}{0.64,0,0.95,0.40}
\definecolor{midgreen}{rgb}{0.0,0.675,0.0}
\definecolor{darkgreen}{rgb}{0,0.5,0}

\usepackage[all]{xy} 
\usepackage{amsfonts}

\begin{document}

\title{Speed of Gravitational Waves and the Fate of Scalar-Tensor Gravity}

\author{Dario Bettoni}
\email{dario.bettoni@nordita.org}
\affiliation{Nordita, KTH Royal Institute of Technology and Stockholm University, \\
Roslagstullsbacken 23, SE-106 91 Stockholm, Sweden}

\author{Jose Mar\'ia Ezquiaga}
\email{jose.ezquiaga@uam.es}
\affiliation{Instituto de F\'isica Te\'orica UAM/CSIC, Universidad Aut\'onoma de Madrid, \\ 
C/ Nicol\'as Cabrera 13-15, Cantoblanco, Madrid 28049, Spain}

\author{Kurt Hinterbichler}
\email{kurt.hinterbichler@case.edu}
\affiliation{CERCA, Department of Physics, Case Western Reserve University, \\
10900 Euclid Ave, Cleveland, OH 44106, USA}

\author{Miguel Zumalac\'arregui}
\email{miguelzuma@berkeley.edu}
\affiliation{Nordita, KTH Royal Institute of Technology and Stockholm University, \\
Roslagstullsbacken 23, SE-106 91 Stockholm, Sweden}
\affiliation{Berkeley Center for Cosmological Physics, LBNL and University of California at Berkeley, \\
Berkeley, California 94720, USA}

\begin{abstract}
The direct detection of gravitational waves (GWs) is an invaluable new tool to probe gravity and the nature of cosmic acceleration. A large class of scalar-tensor theories predict that GWs propagate with velocity different than the speed of light, a difference that can be $\mathcal{O}(1)$ for many models of dark energy.
We determine the  conditions behind the anomalous GW speed, namely that the scalar field spontaneously breaks Lorentz invariance and couples to the metric perturbations via the Weyl tensor. 
If these conditions are realized in nature, the delay between GW and electromagnetic (EM) signals from distant events will run beyond human timescales, making it impossible to measure the speed of GWs using neutron star mergers or other violent events.
We present a robust strategy to exclude or confirm an anomalous speed of GWs using eclipsing binary systems, whose EM phase can be exquisitely determined. he white dwarf binary J0651+2844 is a known example of such system that can be used to probe deviations in the GW speed as small as $c_g/c-1\gtrsim 2\cdot 10^{-12}$ when LISA comes online. This test will either eliminate many contender models for cosmic acceleration or wreck a fundamental pillar of general relativity.
\end{abstract}

\date{\today}



 
\maketitle
 
\paragraph{\bf Introduction and summary.}
 
The direct detection of gravitational radiation \cite{Abbott:2016blz,Abbott:2016nmj} has initiated a new era for astronomy, astrophysics and fundamental physics. The observed gravitational wave (GW) events and the ones to come will usher in novel ways to test the nature of gravity \cite{TheLIGOScientific:2016src}. Here, we will argue that probing the speed of GWs will be a decisive test for gravity and dark energy models.

The nature of the propagation of GWs is a question of great and fundamental interest. Einstein's General Relativity (GR) predicts two massless tensor polarizations each traveling at the speed of light, $c$, with an amplitude inversely proportional to the distance from the source \cite{misner1973gravitation}.  However, major outstanding theoretical issues such as the nature of dark energy and dark matter have led to consider the possibility that gravity differs from GR in some regimes (see e.g. \cite{Clifton:2011jh,Joyce:2014kja} for reviews).  In alternative theories of gravity, additional polarizations may propagate, each with potentially different velocities, attenuations and effective masses \cite{clifford1981theory}. This issue has been well studied in cosmology, and has been a topic of discussion in connection to the early \cite{Amendola:2014wma,Raveri:2014eea,Creminelli:2014wna,DeFelice:2014bma} and the late Universe \cite{Bellini:2014fua,Saltas:2014dha,Lombriser:2015sxa,Zumalacarregui:2016pph}. There are fairly model-independent tests for effects caused by additional polarizations \cite{Will:2014kxa}, damping \cite{Deffayet:2007kf,Calabrese:2016bnu,Garcia-Bellido:2016zmj}, mass \cite{deRham:2016nuf}, and Lorentz symmetry violations \cite{Yagi:2013qpa,Shao:2013wga}.
Up to date, the speed of GWs has been upper bounded with the arrival timing of GW150914 between the two LIGO detectors \cite{Blas:2016qmn}. Also, it has been constrained at the $\sim 1\%$ level from the variation of the orbital period in binary pulsars \cite{Jimenez:2015bwa}. Moreover, if $c_g<c$, a very stringent lower bound $ c_g/c -1\gtrsim -10^{-15}$ can be obtained from the absence of gravitational Cherenkov radiation, probed the observation of ultra-high energy cosmic rays \cite{Caves:1980jn,Moore:2001bv}.

In this paper, we analyze the speed of GWs, $c_{g}$, in generic scalar-tensor theories of gravity and ask when it can differ from the speed of light, $c$. Unlike previous studies, we do not assume a specific cosmological background, instead focusing on the local speed of gravity.  Such anomalous propagation is potentially observable if both gravitational waves and an electromagnetic (EM) or other non-gravitational counterpart signal can be seen from the same source.   

One of two scenarios will arise: the simultaneous arrival of a GW signal with non-gravitational counterpart from a distant source will set extremely stringent and model independent bounds on $c_g$.  However, a very slight difference in propagation speed (as predicted by many models of cosmic acceleration), would cause a delay between the signals' arrival much larger than the multi-messenger observation campaign.  In this case a GW signal never gets identified with its true EM counterpart and other techniques must be used.  We will discuss one such method, the \emph{phase lag test with eclipsing binaries}, based on monitoring periodic galactic sources observable in GW by future space missions such as LISA \cite{AmaroSeoane:2012km}, and in EM by other means, and comparing the phase of the two signals.

A measurement of non-trivial $c_g$ would have profound implications for our understanding of gravity. As we shall see, the anomalous propagation of GW is directly related to fundamental properties of the underlying gravitation theories which can hence be distinguished on this basis.
Conversely, an observation consistent with GWs traveling at the speed of light will place much more severe constraints than any other available test on the large class of theories predicting an anomalous GW speed. In fact, current cosmological constraints on general scalar-tensor theories are only of the order of $\mathcal{O}(1-0.5)$ \cite{Bellini:2015xja}, while future forecasts will reach $\mathcal{O}(0.1-0.01)$ \cite{Alonso:2016suf}. Testing the speed of GWs will dramatically improve these constraints to $\mathcal{O}(10^{-12}-10^{-17})$.

\paragraph{\bf Scalar fields and the speed of GW$\rm s$.}\label{sec:quartic_theory}

In the following, we are going to present a general method to compute the speed of GWs.
 Let us start with an example theory that predicts anomalous GWs propagation: a quartic shift-symmetric Horndeski theory \cite{Horndeski:1974wa,Deffayet:2011gz} $S=\int d^4x\ \sqrt{-g}\mathcal{L}$ with
\begin{eqnarray} \label{eq:ex_theory}
\mathcal{L} =  G(X)R+G'(X)\left((\square\phi)^2-\nabla_\mu\nabla_\nu\phi\nabla^\mu\nabla^\nu\phi\right)\,,
\end{eqnarray}
where $X\equiv-\frac{1}{2}(\partial\phi)^2$ and $G'\equiv\partial G/\partial X$. We set $c=1$ in this section.
Expanding around a background solution,
$ g^{\rm }_{\mu\nu}\rightarrow g_{\mu\nu}+h_{\mu\nu}$, $\phi^{\rm } \rightarrow \phi + \varphi$,
yields a quadratic action for the fluctuations
\begin{equation}
\mathcal{L} = \frac{1}{2}
\left[h_{\mu\nu}{\cal D}^{\mu\nu,\rho\sigma} h_{\rho\sigma}+h_{\mu\nu}{\cal D}^{\mu\nu}\varphi+\varphi {\cal D}\varphi\right],\label{eq:quad_action}
\end{equation}
where ${\cal D}^{(\cdots)}$ represent differential operators depending on the background fields $g_{\mu\nu}$ and $\phi$ and their derivatives.

Since we are interested in local propagation, we adopt Riemann normal coordinates around a point $P$ and expand the scalar and metric background in a Taylor series about $P$,
$g_{\mu\nu} =\eta_{\mu\nu}-{1\over 3}R_{\mu\rho\nu\sigma}x^\rho x^\sigma+\cdots$, 
$\phi=\phi_0+\phi_\mu x^\mu+{1\over 2}\phi_{\mu\nu} x^\mu x^\nu+\cdots $,
where $\phi_\mu=\nabla_\mu\phi$, $\phi_{\mu\nu}=\nabla_\mu\nabla_\nu\phi$ and the derivatives and curvatures are all evaluated at $P$. This leaves freedom for a rotation and boost around P.

We may now zoom in and obtain an effective action valid around the point $P$ by taking the scaling limit, $\lambda\rightarrow0$, with
\begin{eqnarray} 
x^\mu\rightarrow \lambda x^\mu\,,\ \ \varphi \rightarrow {1\over \lambda}\varphi\,,\ \ h_{\mu\nu} \rightarrow {1\over \lambda}h_{\mu\nu}\,. \label{eq:scaling_lim}
\end{eqnarray}
The result is a flat space action, depending on the background field values and derivatives evaluated at $P$.

We will focus on the spin-2 polarizations present in GR and neglect the additional scalar mode.
Imposing the transverse gauge condition $\partial^\mu h_{\mu\nu}=0$, the scaling-limit action reads
\begin{equation} 
\mathcal{L}=\frac{1}{2}  h_{\mu\nu} \left[ G\Box +G^\prime \phi^{\rho}\phi^{\sigma}\partial_\rho\partial_\sigma\right]h^{\mu\nu} 
+ h_{\mu}^{\ \rho} G^\prime \phi^{\mu}\phi^{\nu} \square h_{\nu\rho} +\cdots\,,
\label{eq:flatTTfqa}
\end{equation}
where we omitted terms involving both the trace of the metric and the scalar field.
We then perform a standard $3+1$ split of $h_{\mu\nu}$ and restrict to the transverse-traceless (TT) part of the spatial metric components $h_{ij}$,
\begin{equation}
h_{00}=0,\ \ \ h_{0i}=0,\ \ \ h_{ij}=h_{ij}^{TT}, \ \ \ \varphi=0, \label{gauge}
\end{equation}
with $\partial^j h_{ij}^{TT}=\delta^{ij}h_{ij}^{TT}=0$. 
We will further assume that the spatial shear of the background scalar configuration is negligible.%
\footnote{The precise condition is $\phi_{ii}-\phi_{jj}, \phi_{ij} \ll G^{\prime}/G$ for $(i\neq j)$. This is satisfied in a boosted frame with $\phi_i=0$ whenever $\phi_\mu$ is time-like.}
This assumption simplifies the analysis, ensuring that $h_{ij}^{TT}$ decouple from the other perturbations and allowing us to ignore the terms omitted in Eq. (\ref{eq:flatTTfqa}), which describe the scalar polarization and non-dynamical metric elements.

If the field gradient $\phi_\mu$ is time-like (as expected for a cosmological contribution) we can rotate the coordinates so that $ \phi^{}_\mu =(\dot{\phi},0,0,0),$ for some constant $\dot{\phi}$. Then, the last term of \eqref{eq:flatTTfqa} does not contribute and
\begin{equation} \label{eq:L_tlike_final}
\mathcal{L} = \frac{1}{2} \left\{ \left[ G - G^\prime \dot{\phi}^2 \right] \left(\dot h_{ij}^{TT}\right)^2
- G \left(\vec\nabla h_{ij}^{TT}\right)^2 \right\} \,,
\end{equation}
from which we can read off the propagation speed
\begin{equation} \label{eq:speed_tlike}
{c_g^2}={1 \over 1-{G'\over G}\dot{\phi}^2}\,.  
\end{equation}
In particular, GR corresponds to $G(X)=const.$ and we recover $c_{g}=1$. 

In the case of a space-like field gradient we can boost our reference frame so that the time component vanishes. Decomposing the gradient in components parallel and perpendicular to the GW propagation, $\phi_{i}=\phi^{\parallel}_{i}+\phi^{\perp}_{i}$ we obtain that the velocity of propagation of GWs depends on the direction as
\begin{equation} \label{eq:speed_spacelike}
{c_g^2}=1+\frac{G'\vert\phi_{\parallel}\vert^{2}}{G+G'\vert\phi_{\perp}\vert^{2}}\,.  
\end{equation}
In general the speed is anisotropic (i.e. dependent on the direction), and equal for both the $+$ and $\times$ GW polarizations.

The scaling limit (\ref{eq:scaling_lim}) elliminates all the lower derivative terms, which is the reason that the resulting GW speed is frequency independent. This is different for other well studied cases, such as massive gravitons \cite{deRham:2010kj} (see \cite{Hinterbichler:2011tt,deRham:2014zqa} for reviews) or Lorentz violations. These other scenarios modify the waveform in a frequency dependent way and can thus be constrained from GW observations alone  \cite{Yunes:2009ke,TheLIGOScientific:2016src,Yunes:2016jcc}.
For the sake of simplicity we have also neglected the scalar mode, which may also have its own anomalous propagation speed \cite{Babichev:2007dw,Sawicki:2015zya,Lindroos:2015fdt,Hagala:2016fks}.

\paragraph{\bf Conditions for anomalous GWs speed.}

We now study the origin of the anomalous speed of GWs (\ref{eq:speed_tlike}, \ref{eq:speed_spacelike}) in more generality. The Lagrangian for the transverse-traceless components (\ref{eq:L_tlike_final}) can be written in terms of an \emph{effective gravitational metric}
\begin{equation}
 \mathcal{L}\propto h_{\alpha\beta}^{TT} \big(\mathcal{G}_{\mu\nu} \partial^{\mu}\partial^{\nu}\big) h^{\alpha\beta}_{TT}\,,
\label{eq:effG}
\end{equation}
determining the causal structure of GW propagation.%
\footnote{We focus on the spin-2 components and assume they decouple. Nonetheless, Eq. (\ref{eq:effG}) remains valid for the propagation eigenstates of the linearized fields (including the scalar mode and the generalization of $h_{\alpha\beta}^{TT}$ when it couples to other perturbations), with a different $\mathcal{G}_{\mu\nu}^{A}$ for each polarization $A$. } 
The propagation path for GWs will be given by the condition $\mathcal{G}_{\mu\nu}dx^\mu dx^\nu=0$ and will in general be different from the lightcone condition $g_{\mu\nu}dx^\mu dx^\nu =0$ unless the two metrics obey a \emph{conformal relation}:  $\mathcal{G}_{\mu\nu}=\Omega(x)g_{\mu\nu}$. The lack of proportionality is found already in the simple example theory (\ref{eq:ex_theory}), where
\begin{equation}\label{eq:eff_G_ex}
\mathcal{G}_{\mu\nu} = G(X)g_{\mu\nu} + G^\prime (X)\phi_\mu \phi_\nu\,, 
\end{equation}
and $\mathcal{G}_{\mu\nu}$, $g_{\mu\nu}$ are connected by a \emph{disformal relation} \cite{Bekenstein:1992pj} for which $\mathcal{G}_{\mu\nu}\neq \Omega(x)g_{\mu\nu}$. Such a relation is ubiquitous in modern scalar-tensor theories \cite{Bettoni:2013diz,Zumalacarregui:2013pma,Gleyzes:2014qga,DAmico:2016ntq}.

Let us examine the conditions for a disformal relation to arise in a generic theory of gravity. First, it is necessary that the background scalar field has a non-trivial configuration that spontaneously breaks Lorentz invariance, e.g. $\phi_\mu\neq 0$ in Eq. (\ref{eq:eff_G_ex}). In addition, we note that the effective second-order Lagrangian (\ref{eq:quad_action}) follows from the second variation of the action over a background, and is hence equal to the first variation of the equations of motion (EoM). The simplest term in the EoM producing second derivatives and entering in (\ref{eq:effG}) is the Ricci curvature. When expanded to first-order, considering only the TT components,
\begin{equation} \label{eq:Ricci}
 R_{\mu\nu}^{TT} = -\frac{1}{2}\Box h_{\mu\nu}^{TT}\,\quad\mathrm{and} \quad R^{TT} = 0\,,
\end{equation}
only contribute to the conformal part in the effective gravitational metric (\ref{eq:effG}). 

Further second derivative terms are restricted by covariance to originate either from the Riemann tensor or repeated application of covariant derivatives (e.g. third derivatives of the scalar field), with the two cases related by $\nabla_\mu \nabla_\nu\phi^\alpha = \nabla_\nu \nabla_\mu\phi^\alpha + R^\alpha_{\;\lambda\mu\nu}\phi^\lambda$. To first-order the TT contribution to the Riemann tensor reads
\begin{equation} \label{eq:Riemann}
R_{\mu\alpha\nu\beta}^{TT} = -\frac{1}{2}\partial_{\beta}\partial_{\alpha} h_{\mu\nu}^{TT}
 +\frac{1}{2}\partial_{\nu}\partial_{\alpha} h_{\mu\beta}^{TT} -(\alpha\leftrightarrow\mu)\,,
\end{equation}
The above expression explicitly induces disformal terms in Eq. (\ref{eq:effG}) via contractions with scalar field derivatives. In the simple example (\ref{eq:ex_theory}), only $\phi^\mu$ enters in the effective metric (\ref{eq:eff_G_ex}) due to the particular non-minimal coupling to the Ricci scalar. In more general cases, for instance when there are couplings to the Ricci tensor such as in quintic Horndeski, second derivatives $\phi^{\mu\nu}$ could appear contracted with the derivatives of the metric and hence in $\mathcal{G}_{\mu\nu}$. Thus, the effective metric would belong to the extended disformal class \cite{Zumalacarregui:2013pma,Ezquiaga:2017ner}. In any case, because the Ricci tensor only contributes to the conformal part, the contribution of $R_{\mu\nu\alpha\beta}$ leading to the anomalous speed of GWs is fully captured by the Weyl tensor (i.e. the trace-free part of the Riemann tensor). 
For the simple theory (\ref{eq:ex_theory}), the Weyl tensor appears explicitly in the equations of motion whenever $G^\prime\neq 0$ \cite{Bettoni:2015wta}.

These considerations allow us to formulate a \textit{Weyl criterion} for anomalous speed of spin-2 GWs. The effective gravitational metric of the example theory (\ref{eq:eff_G_ex}) can be generalized to 
\begin{equation}\label{eq:effG2}
 \mathcal{L}\propto h_{\mu\nu}\lp \C\Box+\W^{(\alpha\beta)}\partial_{\alpha}\partial_{\beta}\rp h^{\mu\nu}\,,
\end{equation}
where $\C$ and $\W^{\mu\nu}$ are the contributions associated with the Ricci and Weyl tensors respectively. Anomalous GW speed requires that $\mathcal{W}^{\alpha\beta}\neq0$, i.e. for the background scalar derivatives to couple to the Riemann/Weyl curvature. If the Weyl factor is purely time-like and constant around P, $\W^{\mu\nu}=\W^{00}\delta^{\mu}_{0}\delta^{\nu}_{0}$, the speed of tensors becomes
\begin{equation}
c_{g}^{2}=\frac{\C}{\C-\W^{00}}\,.
\end{equation}

In Horndeski theories, which is a general framework that englobes most of the current dark energy models, the EoM are second order \cite{Horndeski:1974wa}. Therefore, the ocurrence of the Weyl tensor fully distinguishes theories in which $c_g=c$ exactly and those in which the speed of GWs is allowed to vary. 
GR, Kinetic Gravity Braiding \cite{Deffayet:2010qz} and Jordan-Brans-Dicke theories \cite{Brans:1961sx} (including $f(R)$ \cite{Sotiriou:2008rp,DeFelice:2010aj}) only contain Ricci curvature in their equations of motion, and therefore do not modify the speed of GWs. On the other hand covariant Galileons \cite{Deffayet:2009wt} and the covariantization of other generalizations \cite{Hinterbichler:2010xn,Goon:2011qf,Goon:2011uw,Trodden:2011xh} will generically predict $c_g\neq c$ \cite{Brax:2015dma}.

Although the Weyl criterion is characteristic of ST theories, the occurrence of a disformal relation can be applied to more general theories such as massive gravity \cite{deRham:2010kj}. In this case the kinetic term has the Einstein-Hilbert form and hence $c_g=c$ plus corrections $\mathcal{O}\left(\frac{m^2}{E^2}\right)$ beyond the scaling limit (\ref{eq:scaling_lim}), as expected from unbroken Lorentz invariance. In the case of bigravity \cite{Hassan:2011zd} the situation is more subtle, as the kinetic term of the second metric $\sqrt{-f}R[f_{\mu\nu}]$ forces its excitations to propagate along $f_{\mu\nu}dx^\mu dx^\nu = 0$, with $f_{\mu\nu}\neq \Omega(x) g_{\mu\nu}$ in non-flat background space-times. Although matter does not couple to $f_{\mu\nu}$ directly the anomalous speed may be detectable via graviton oscillations \cite{Narikawa:2014fua}. Many theories that attempt to explain away dark matter such as TeVeS also predict an anomalous GW speed \cite{Desai:2008vj}.

\paragraph{\bf Phase lag test with eclipsing binaries}

Most of the present bounds on $c_{g}$ can be significantly strengthened by comparing GWs with other signals.
In theories in which matter is universally coupled to the metric, electromagnetic signals and ultrarelativistic particles propagate at the speed of light. 
This produces a delay between GW and electromagnetic signals
\begin{equation}\label{eq:t_difference}
\Delta t
= r\left(\frac{1}{c_g} - \frac{1}{c} \right)
\equiv \frac{r}{c} \varepsilon_g \approx  10^{14} s \frac{r}{\rm Mpc} \varepsilon_g \,,  
\end{equation}
where we define the \emph{differential delay parameter} $\varepsilon_g\equiv c\partial\Delta t/\partial r$ (in general space-times $r$ is the proper distance and one has to correct for time dilation at emission \cite{Will:2014kxa}).
The detection of violent, multi-messenger events at cosmological distances bears the promise of phenomenal constraints, even in the presence of considerable astrophysical uncertainties. LIGO expects to perform such measurements using violent events such as binary compact object mergers involving neutron stars \cite{lrr-2016-1}.

However, no distant GW-EM event will possibly be observed if $c_g$ is modified significantly, since the delay between both signals will be much larger than the monitoring time around the GW detection.
This is the case of cosmic acceleration models without a cosmological constant such as covariant Galileons \cite{Deffayet:2009wt,Barreira:2014jha}, for which $|c_g/c -1|\sim 10-100\%$ (see \cite{Brax:2015dma} and Fig. 1 of Ref. \cite{Renk:2016olm}). 
If such a model is responsible for cosmic acceleration, the arrival times of both signals will differ by millions or even billions of years.
Clearly, an alternative test for the speed of GWs would be needed in this situation. In the following, we discuss how observations of sources with periodic signals can help to test whether $c_g=c$. In particular, we propose a \emph{phase lag test with eclipsing binaries} that overcomes this limitation.

The anomalous speed of GWs can be tested by monitoring periodic sources with both GW and EM emission \cite{Larson:1999kg,Cutler:2002ef}.
This ensures that both signals can be observed continuously and allows for a long observation period.
A suitable source is a binary system in the band of space-based interferometers \cite{lrr-2013-7}, including \emph{verification binaries} \cite{Stroeer:2006rx,AmaroSeoane:2012je,28097}: systems expected to be resolvable by LISA and which have already been identified and characterized using electromagnetic observations (see Ref. \cite{28097} for an updated list). 
An extraordinarily clean binary system is WDS J0651+2844: a binary, detached white dwarf system $\sim 1$kpc away from the Sun and whose orbital plane is approximately aligned with the Solar System, allowing the observation of periodic eclipses \cite{Brown:2011gq}. Its short orbital period $\sim 12.75$ min falls within the LISA band and makes it a loud GW source, in which the effect of GW emission has already been observed by the period variation \cite{Hermes:2012us}.

\begin{figure}[t!]
 \includegraphics[width=\columnwidth]{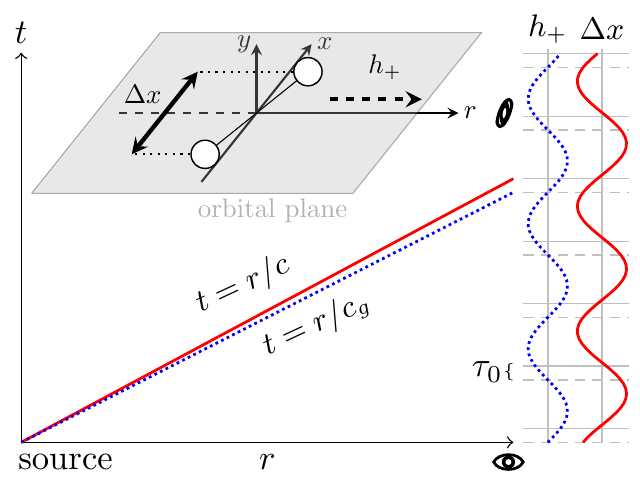}
 \caption{The phase lag test for the speed of gravity. A compact binary system such as WDS J0651+2844 is monitored both electromagnetically and using GWs. For this geometry (top) only the $+$ GW polarization is emitted in the observer's direction. Its amplitude $h_+$ is initially correlated with the object transverse separation $\Delta x$, but a phase lag (\ref{eq:phase_lag})  accumulates on the propagation if $c_g \neq c$ (bottom and right).
}\label{fig:pulsar} 
\end{figure}

Let us model WDS J0651+2844 as a binary orbit coplanar with the observer and at a distance $r$ from it, cf Fig. \ref{fig:pulsar}. Due to symmetry the gravitational radiation emitted in the observer's direction will be predominantly in the $+$ polarization $h_{ij}=h_+(t)(\hat x\hat x - \hat y \hat y)$.%
\footnote{The orbital inclination is $\iota = 86.9^{+1.6}_{-1.0}$ degrees  \cite{Brown:2011gq}, making $h_\times$ suppressed by $\cos(\iota)\approx 0.05$ in amplitude and shifted $\pi/2$ in phase relative to the $+$ component.}
Assuming GR (i.e. $c_g=c$), the $h_+$ polarization will be in phase with $\Delta x$, the distance between the objects transverse to the line of sight as observed electromagnetically.  Therefore, although the components of the binary will not be resolvable, $\Delta x = 0$ coincides with the eclipses and can be timed with extraordinary precision \cite{Larson:1999kg}.

In theories other than GR, the EM and GW observables will evolve as periodic functions of different retarded times, i.e. $ \Delta x \propto \cos(2\omega(t-r/c))$ and $h_+\propto \cos(2\omega(t-r/c_{g}))$.
The difference in propagation speed accumulated over the propagation distance $r$ produces a \emph{phase lag} between the GW and the EM signals:%
\footnote{We have neglected the delay from the atmospheric or interstellar refractive index, which can be shown to be unimportant \cite{Larson:1999kg}.}
\begin{eqnarray} \label{eq:phase_lag}
\Delta \Phi(t) &=& 2\omega \frac{r(t)}{c}\left(\frac{c}{c_g} - 1 \right) 
= 2\omega \frac{r(t)}{c}\varepsilon_g \,,
\end{eqnarray}
where the distance between source and detector
\begin{equation}\label{eq:distance}
r(t) = r_0 + v_{\rm rel}\, t + r_{\rm orb}(t)\,, 
\end{equation}
includes the initial separation, relative velocity and the detector's orbit. We will focus on the effect of $r_0,v_{\rm rel}$, as the effect of $r_{\rm orb}$ has been considered \cite{Finn:2013toa}.

For eclipsing binaries we can neglect the error in EM measurements in constructing the relative phase (\ref{eq:phase_lag}) $\Delta\Phi(t)\equiv 2\omega(\tau_0+ \hat \beta t)$. The precision will be then limited by our knowledge of the GW signal. We can obtain an estimate of fo the  1-$\sigma$ uncertainties using the Fisher matrix formalism \cite{Vallisneri:2007ev} for the following quantities:
\begin{eqnarray}
\tau_0 \equiv  \varepsilon_g \frac{r_0}{c}
\,,\;
\Delta \tau_0 = \frac{1}{\sqrt{2}\, \omega\, \Sigma} 
\approx 0.2 s \left(\frac{2\pi/\omega}{765s}\right)\left(\frac{T}{5y}\right), 
\label{eq:Delta_tau}
\\
\hat \beta \equiv  \varepsilon_g \frac{v_{\rm rel}}{c}
\,,\;
\Delta \hat \beta = \frac{\sqrt{3/2}}{\omega T \Sigma}
\approx 10^{-8} \left(\frac{2\pi/\omega}{765s}\right)\left(\frac{T}{5y}\right)
\label{eq:Delta_beta}
, 
\end{eqnarray}
where $T$ is the observation time and $\Sigma$ denotes the total signal-to-noise ratio of the GW detection (see Appendix). The expected detection significance of verification binaries with LISA is $\Sigma\sim 100 \left(\frac{T}{1y}\right)$ \cite{Finn:2013toa}. 

A non-zero measurement of either (\ref{eq:Delta_tau}, \ref{eq:Delta_beta}) represents a smoking gun for $c_g\neq c$:
\begin{itemize}
\item \textit{$\tau_0$:}  The relative phase of the signals can detect an anomalous propagation speed in the range 
$|\varepsilon_g| \gtrsim 2\cdot 10^{-12}\left(\frac{\text{kpc}}{r_0}\right)\Big(\frac{\Delta\tau_0}{0.2s}\Big)$. The false-negative case where $ 2 r_0 \varepsilon_g \omega/(c\pi) $ equals an integer within the measurement error is very unlikely (${\rm prob.}\approx \Sigma^{-1} \sim 0.2\%$) and can be excluded by observing multiple systems or measuring the frequency shift $\hat\beta$. 

\item \textit{$\hat\beta$:} The relative velocity of the system induces a frequency shift, sensitive to anomalous GW speeds in the range 
$|\varepsilon_g|\gtrsim 10^{-4}\left(\frac{30{\rm km}/s}{v_{\rm rel}}\right)\left(\frac{\Delta\beta}{10^{-8}}\right) $.
Despite the $(\omega T)^{-1}$ gain when observing over many cycles, this test is less competitive due to the non-relativistic factor.
\end{itemize}
Note that both the measurement of the relative phase and the velocity can be used as a test of $\varepsilon_g\neq0$ \emph{and} as a measurement of $c_g$. The latter application requires a measurement of either $r_0$ or $v_{\rm rel}$, which will almost certainly dominate the error . Nevertheless, clean systems such as WDS J0651+2844 will be able to confirm deviations from $c_g = c$ at the level of few parts in a trillion.

\paragraph{\bf Conclusions.}
Many well studied models of dark energy and modified gravity theories predict an anomalous local speed of gravity around non-trivial backgrounds.
The \emph{Weyl criterion} provides a clear-cut way to distinguish two classes of gravitational theories, those for which the speed of GWs is \emph{exactly equal} to the speed of light, and those in which it can vary depending on the theory parameters and the background configuration of the scalar field. 
Future multi-messenger GW observations will probe this effect to exquisite precision: if the prediction of GR is satisfied, this will place such stringent constraint on theories allowing variations in the speed of GWs, $\mathcal{O}(10^{-17})$, that they will become uninteresting for any low energy application, including cosmic acceleration. 
On the other hand, a confirmation of an anomalous propagation of GWs by extragalactic and galactic sources would be able to rule out GR and all other theories with simple kinetic terms, which would significantly impact our understanding of gravity.
This could be achieved applying the proposed phase lag test for eclipsing binaries to the already identified white dwarf binary WDS J0651+2844. Any of these two scenarios shows that the speed of GWs will be by far one of the most powerful tools to constrain gravity and dark energy models.

\begin{acknowledgments}
Acknowledgements: We are grateful to J. Beltran-Jimenez, D. Blas, S. Cespedes, D. G. Figueroa, J. Garcia-Bellido, L. Hui, B. Kanniah, I. Sawicki, and N. Wex for useful conversations. 
DB acknowledges financial support from "Fondazione Angelo Della Riccia".
JME is supported by the Spanish FPU Grant No. FPU14/01618 and thanks  Nordita for hospitality. 
\end{acknowledgments}

\bibliography{gw_refs}

\appendix

\section{Signal to noise estimates}
\label{app:signal_to_noise}

The signal-to-noise ratio $\Sigma$ for a GW detection is given by
\begin{equation}\label{eq:signal_def}
 \Sigma^2  = \frac{1}{\sigma_f^2}\int_0^T \tilde R^2(t) dt \equiv \varrho\,.
\end{equation}
Here $\tilde R$ is the response of the detector to the signal and $\sigma_f^2$ is the noise power at the GW frequency. We assume the GW to be monochromatic and follow Ref. \cite{Finn:2013toa} (see Ref. \cite{Vallisneri:2007ev} for further details and cautionary notes). For a given detector the response function depends on the GW polarizations as $\tilde R(t) = A_+(t)h_+ + A_\times (t) h_\times $ where $A_i$ contain information of the antenna pattern of the detector and its orientation as a function of time. However, as discussed in the text, we will consider the situation in which only one polarization is received and assume that the errors in the electromagnetic signal are negligible. Therefore we can reconstruct the relative phase (Eq. \ref{eq:phase_lag} in main text) directly
\begin{equation}
 \tilde R(t) = \varUpsilon\cos(\varpi t + \psi)\,,
\end{equation}
where the signal has an overall amplitude $\varUpsilon$, which will not directly affect the reconstruction of $\psi$ and $\omega$.

The Fisher matrix is then given as the derivative of Eq. (\ref{eq:signal_def}) with respect to the model parameters
\begin{equation}
 F_{ij}=\frac{2}{\sigma_f^2} \int_0^T \frac{\partial \tilde R}{\partial \theta_i} \frac{\partial \tilde R}{\partial \theta_j}dt\,,
\end{equation}
where $\theta_i=(\varUpsilon,\varpi,\psi)$ collectively denotes the unknown parameters of the signal. The error in the parameter $\theta_i$ assuming the other ones are perfectly known is $(F_{ii})^{-1/2}$, while the error in a parameter marginalized over the rest is $\sqrt{(F^{-1})_{ii}}$.

The Fisher matrix elements read
\begin{eqnarray*}
 F_{\varUpsilon\varUpsilon} &=& \frac{2}{\sigma_f^2}\int \cos^2(\varpi t+\psi)dt 
=  2 \varrho/\varUpsilon^2\,,\\
 F_{\varUpsilon\varpi} &=& \frac{2}{\sigma_f^2}\int - t\sin(\varpi t+\psi) \varUpsilon \cos(\varpi t+\psi)dt 
 \sim \text{osc.}\,,  
  \end{eqnarray*}
 \begin{eqnarray*}
 F_{\varUpsilon\psi} &=& \frac{2}{\sigma_f^2}\int - \varUpsilon \cos(\varpi t+\psi)\sin(\varpi t+\psi)dt \sim  \text{osc.}\,,  \\
 F_{\varpi\varpi} &=& \frac{2}{\sigma_f^2}\int \varUpsilon^2 t^2\sin^2(\varpi t+\psi)dt 
 =  2 \varrho \frac{t^2}{3}  +  \text{osc.}\,, \\
  F_{\varpi\psi} &=& \frac{2}{\sigma_f^2}\int \varUpsilon^2 t \sin^2(\varpi t+\psi)dt 
 =   \varrho {t} +  \text{osc.}\,, \\
 F_{\psi\psi} &=& \frac{2}{\sigma_f^2}\int \varUpsilon^2 \sin^2(\varpi t+\psi)dt 
 =  2 \varrho  +  \text{osc.}\,,
\end{eqnarray*}
where osc. denotes oscillatory terms that become negligible for $T\gg \varpi^{-1}$ and we have used $\varrho= {\varUpsilon^2\over 2\sigma_f^2}T$.
Since $F_{\varUpsilon\varpi},\,F_{\varUpsilon\psi}$ do not build up with time, the amplitude is uncorrelated with the frequency and the phase. However, $\varpi,\psi$ are correlated with one another. The Fisher matrix and its inverse for the $(\varpi,\psi)$ subspace are 
\begin{equation}
 \hat{F} = \varrho 
 \left(
\begin{array}{cc}
 \frac{2}{3}T^2 & {T} \\
 {T} & 2 \\
\end{array}
\right)\,, \qquad 
 \hat{F}^{-1} = \frac{1}{\varrho}
 \left(
\begin{array}{cc}
 \frac{6}{T^2} & -\frac{3}{T} \\
 -\frac{3}{T} & 2 \\
\end{array}
\right)\,.
\end{equation}
From which we read the errors in the phase and frequency
\begin{eqnarray}
 \Delta \psi = \frac{\sqrt{2}}{\Sigma}\,,
 \quad
 \Delta \varpi = \frac{\sqrt{6}}{T \cdot \Sigma}\,,
\end{eqnarray}
which translate straightforwardly into the results (Eqs. (\ref{eq:Delta_tau}, \ref{eq:Delta_beta}) in the main text).

\end{document}